\documentclass{aa}
\usepackage{graphicx,times}
\newlength{\fsize}
\def\Real{{\rm I\mathchoice{\kern-0.70mm}{\kern-0.70mm}{\kern-0.65mm}%
  {\kern-0.50mm}R}}  
  \def\bx#1{\leavevmode\thinspace\hbox{\vrule\vtop{\vbox{\hrule\kern1pt
  \hbox{\vphantom{\tt/}\thinspace{\bf#1}\thinspace}}
  \kern1pt\hrule}\vrule}\thinspace}

\def\be{\begin{equation}} \def\ee{\end{equation}}

\begin{document}

\title{A differential rotation driven dynamo in a stably stratified star}

\author{J.\ Braithwaite and H.C.\ Spruit}
\institute{Max-Planck-Institut f\"ur Astrophysik, Karl-Schwarzschild-Stra{\ss}e 1,
Postfach 1317,  D--85741 Garching, Germany}
\authorrunning{J. Braithwaite, H. Spruit}
\offprints{henk@mpa-garching.mpg.de} \date{Received / Accepted}
\titlerunning{A  dynamo in a stably stratified star}

\abstract{We present numerical simulations of a self-sustaining
magnetic field in a differentially rotating non-convective stellar
interior. A weak initial field is wound up by the differential
rotation; the resulting azimuthal field becomes unstable and produces 
a new meridional field component, which is then wound up anew, thus
completing the `dynamo loop'. This effect is observed both with and
without a stable stratification. A self-sustained field is actually  obtained 
more easily in the presence of a stable stratification. The results
confirm the analytical expectations of the role of Tayler instability.
\keywords {magnetohydrodynamics (MHD) -- stars:
magnetic fields -- stars: differential rotation}}

\maketitle

\section{Introduction}
\label{sec:intro}

It has long been known that magnetic fields can be generated by a
dynamo operating in the {\it convective} zone of a differentially
rotating star (e.g. \cite{Parker:1979}). A toroidal field is produced
by winding-up of the poloidal (meridional) component, and the bubbles
of gas moving upwards and downwards move perpendicular to these 
toroidal field lines, bending them and creating a new poloidal component,
closing the `dynamo loop'. This type of dynamo has been the subject of
extensive work over several decades. A process like this has been held 
responsible for the magnetism of stars with convective envelopes like the Sun. 

Whether
or not such direct causation by convection is actually the correct explanation
of dynamos like the solar cycle\footnote{It is in fact more likely that buoyant
instability of the magnetic field, rather than convection, is the key process
in the solar cycle, cf. the discussion in Spruit (1999).}, the predominance of this view 
has obscured the fact that self-sustained magnetic fields do not require the 
presence of convection or other imposed small scale velocity fields. A 
magnetic field can produce small-scale perturbations from its own 
instability, without recourse to externally imposed perturbations. A 
well-known example in the context of accretion discs, where a dynamo was 
produced when differential rotation wound up a field which was then subject 
to magnetohydrodynamic instability (Hawley et al. 1996). 

The same
principle was applied by Spruit (2002) to differentially rotating stars. 
In this scenario, a toroidal field is wound up by differential rotation from 
a weak seed field. The field remains predominantly toroidal, subject to 
decay by instability of the field, but is continuously regenerated by the 
winding-up of  irregularities produced by the instability. This scenario has 
been applied in stellar evolution calculations of the internal rotation of 
massive stars by Heger et al. (2003) and Maeder \& Meynet (2003).
The process is conceptually similar to the small-scale self-sustained fields
found in MHD simulations of accretion discs, but operates on a different
form of MHD instability (pinch-type or Tayler instability as opposed
to magnetorotational, cf. Spruit 1999).

A self-sustained field of this type could have important implications
for not only the magnetism of a star, but also for the transfer of
angular momentum. Differential rotation is created, when the star is 
formed, as a consequence of conservation of angular momentum when 
parts of it contract or expand, and through angular momentum loss through
a stellar wind (`magnetic braking'). In the absence of a magnetic field,
kinetic viscosity would eventually damp differential rotation, but
only on a time-scale much longer than the lifetime of the star. If a
weak magnetic field were present in a star with infinite conductivity,
the field would be wound up, its Lorentz force exerting a force back
on the gas, tending to slow the differential rotation. If we assume
that no magnetohydrodynamic instabilities were present, the energy of
the field would become eventually comparable to the kinetic energy of 
the differential rotation, and the field would exert a force on the gas 
strong enough to reverse the differential rotation. Oscillations would 
follow, with energy continuously being transferred from kinetic to 
magnetic and back again (\cite{Mestel:1953}). 
Finite conductivity would have the effect of damping these
oscillations. However, if the magnetic field became unstable, as we
expect it to, the energy of the field need never reach a level
comparable to the kinetic energy and the direction of differential
rotation would never be reversed. Instead, differential rotation would
gradually be slowed, and the magnetic field held at a low steady-state
level. This could be what has happened in the radiative core of the Sun, 
explaining the near-uniform rotation there
(\cite{Schouetal:1998}, \cite{Charbonneauetal:1999}).

While such a dynamo process is plausible, it has so far been described only in 
terms of an elementary scaling model (Spruit 2002). With the calculations presented
here we verify, first of all, the existence of a self-sustained field generation 
process. At the next level, the goal is to compare the properties of the numerical 
results with the predictions of the model, in particular concerning the central 
role of Tayler instability. The field strength resulting from the dynamo 
process depends on the balance between the decay of the toroidal field and 
the winding up of irregularities. The analytic model only presents the basic scaling of this
balance with parameters like the strength of the differential rotation; the actual
values of the coefficients in this scaling have to be found by numerical simulations.

\subsection{Instability of a toroidal field}
\label{sec:inst}
Tayler (1973) and Acheson (1978)
looked at toroidal fields in stars, that is, fields that have only an
azimuthal component $B_\phi$ in a cylindrical coordinate frame $(\varpi,\phi,z)$
with the origin at the centre of the star.
With the energy method, Tayler derived necessary and sufficient stability conditions
in adiabatic conditions (no viscosity, thermal diffusion or magnetic diffusion). 
The main conclusion was that such purely toroidal fields are always
unstable at some place in the star, in particular to perturbations of the $m=1$ form, 
and that stability at any particular place does not depend on field strength but 
only on the geometry of the field configuration. 
An important corollary of the results of Tayler (1973, esp. the Appendix) was
the proof that instability is {\it local} in meridional planes. If the necessary 
and sufficient condition for instability is satisfied at any point $(\varpi,z)$,
there is an unstable eigenfunction that will fit inside an infinitesimal environment
of this point. The instability is always global in the azimuthal direction, however. 
The instability takes place in the form of a low-azimuthal order displacement in 
a ring around the star. Connected with this is the fact that the growth time of
the instability is of the order of the time it takes an Alfv\'en wave to travel 
around the star on a field line. This and other instabilities were reviewed by 
Spruit (1999).

In Braithwaite (2005) we investigated the development of this instability
numerically, the results confirming the conclusions from the previous analytical
work. We showed that a toroidal field of strength $B=B_0 \varpi/\varpi_0$ (where 
$B_0$ and $\varpi_0$ are constants) in a stably stratified atmosphere is unstable 
on the axis to perturbations of the $m=1$ type.  We confirmed that the growth rate
$\sigma$ of the instability is approximately equal to the local Alfv\'en frequency
$\omega_\mathrm{A}$, given by:
\begin{equation}
\omega_\mathrm{A}\equiv\frac{v_\mathrm{A}}{\varpi}=\frac{B}{\varpi\sqrt{4\pi \rho}}=\frac{B_0}{\varpi_0\sqrt{4\pi \rho}},
\label{eq:defofoa}
\end{equation}
where $v_\mathrm{A}$ is the Alfv\'en speed and $\rho$ is the density. We also 
confirmed that rotation about an axis parallel to the magnetic axis can suppress 
the instability if $\Omega>\omega_A$. However, this stabilization only takes place 
for the case with neither thermal nor magnetic diffusion ($\kappa=\eta=0$). It is 
not entirely certain what effect rotation may have when these two are
present, although it seems very likely that in the limit $\Omega\gg
\omega_\mathrm{A}$, the growth rate is merely reduced by a factor
$\omega_\mathrm{A}/\Omega$, so that:
\begin{eqnarray}
\sigma\approx\omega_\mathrm{A} \,& &  \qquad\qquad ( \Omega
\ll \omega_\mathrm{A} ),
\label{eq:sigma_low_o}\\
\sigma\approx\frac{\omega^2_\mathrm{A}}{\Omega} & & \qquad\qquad ( \Omega
\gg \omega_\mathrm{A} ).
\label{eq:sigma_high_o}
\end{eqnarray}

In the unstratified case where the magnetic diffusivity is zero ($\eta=0$), 
all vertical wavelengths are unstable. However, stratification stabilizes the 
longest vertical length scales. This is because it discourages any vertical 
motion, which is greatest in modes of large vertical scale. Magnetic diffusion 
stifles the shortest wavelengths, since fluctuations in the magnetic field 
produced by the instability are smoothed out by the diffusion at a rate which
depends on the length scale of the fluctuations. If $n$ is the
vertical wavenumber of the unstable mode, then
\begin{equation}
\frac{\sigma}{\eta} > n^2 > \frac{N^2}{\omega_{\rm A}^2 \varpi_0^2},
\label{eq:minandmaxofn}
\end{equation}
where $\varpi_0$ is some measure of the extent of the field in the horizontal 
direction.

The two effects can conspire to kill the instability completely when the
upper and lower limits on wavelength meet each other. This puts a lower
limit on the field strength for instability, expressed by the inequalities:
\begin{eqnarray}
\omega_{\rm A}^3 > \frac{\eta N^2}{\varpi_0^2} & &  \qquad\qquad ( \Omega
\ll \omega_\mathrm{A} ),
\label{eq:nin_low_o}\\
\frac{\omega_{\rm A}^4}{\Omega} > \frac{\eta N^2}{\varpi_0^2} & &\qquad\qquad 
( \Omega \gg \omega_\mathrm{A} ).
\label{eq:nin_high_o}
\end{eqnarray}
For the core of the present Sun, this yields a minimum field strength
of the order $3\times 10^3$ G. 

\subsection{Expected properties of the dynamo}
\label{sec:properties}

In this section we summarize the scenario of Spruit (2002)  for the generation 
of a self-sustained magnetic by differential rotation in a stably stratification.

A weak initial field is wound up by differential rotation. After only a few 
differential turns the field is predominantly toroidal ($B_\phi\gg B_\varpi$ 
and $B_\phi\gg B_z$). Eventually the field strength at which instability sets 
in will be reached (given by Eqs.~(\ref{eq:nin_low_o}) and (\ref{eq:nin_high_o})). 
There are various time-scales of relevance here, the shortest being the
reciprocal of the Brunt-V\"ais\"al\"a (buoyancy) frequency $N$,
given by $N^2=(g/T)({\rm d}T/{\rm d}z + g/c_{\rm p})$.
Provided the star is rotating at less than the break-up rate, the
rotational frequency $\Omega$ will be smaller than $N$. In most
stars, $\Omega$ will however be greater than the magnetic frequency
$\omega_{\rm A}$ (see Eq.~(\ref{eq:defofoa})). The differential
rotation time-scale is given by
\begin{equation}
\tau_{\rm dr}=(\varpi_0 \partial_z \Omega)^{-1}.
\label{eq:tau_dr}
\end{equation}
It is this time-scale $\tau_{\rm dr}$ which will determine how quickly 
the initial field is wound up into a predominantly toroidal field.

Spruit (2002) derives properties of the dynamo in the case where:
\begin{equation}
N \gg \Omega \gg \omega_{\rm A}.
\label{eq:henksordering}
\end{equation}
This is the most realistic regime. In addition, we expect in a real
star to have $\tau_{\rm dr}$ of the same order as, but in most regions
probably greater than, $\Omega^{-1}$.

At the time when the instability sets in, the growth time of the
instability is so long (from Eqs.~(\ref{eq:sigma_low_o}) and
(\ref{eq:sigma_high_o})) that the field is still being wound up
faster than it is able to decay. However, as the field grows,
and the instability growth rate $\sigma$ rises, a point will be
reached where the field is decaying and being wound up at the same
rate -- we call this `saturation'. The time-scale $\tau_{\rm a}$ on
which the field component $B_z$ is wound up into an azimuthal
component of comparable strength to the existing azimuthal field is
given by
\begin{equation}
\tau_{\rm a} = \tau_{\rm dr}\frac{B_\phi}{B_z}.
\label{eq:tau_a}
\end{equation}
Thus, the greater $B_z$, the shorter the amplification
time-scale. At this point, we need to know the value of $B_z$. 
This is provided by the instability which produces a vertical component
from the azimuthal field by the unstable fluid displacements. If these have
a vertical length scale $l$ and horizontal scale $\varpi_0$, we have
\begin{equation}
B_z \approx B_\phi l/\varpi_0,
\label{eq:b_z}
\end{equation}
So, $B_z$ is greatest for the unstable modes with greatest vertical 
wavelength, i.e. with lowest wavenumber $n=1/l$. If we equate this minimum
amplification time-scale to the instability time-scale $\sigma^{-1}$,
we get, using Eqs.~(\ref{eq:minandmaxofn}) and (\ref{eq:b_z}),
\begin{equation}
\omega_{\rm A} \approx \frac{\Omega}{N\tau_{\rm dr}},
\label{eq:high_o_saturation}
\end{equation}
in the case where $\Omega \gg \omega_\mathrm{A}$. In the slowly
rotating ($\Omega \ll \omega_\mathrm{A}$) case, we find that
$\omega_{\rm A}$ drops out of the equation leaving $N=\tau_{\rm
dr}^{-1}$3. In this case, therefore,
the field will either grow until it is strong enough to kill the
differential rotation itself, or it will decay until the $\Omega \gg
\omega_\mathrm{A}$ regime is reached. If $\tau_a < \sigma^{-1}$, the
field will grow, i.e. if
\begin{equation}
N < \tau_{\rm dr}^{-1}.
\label{eq:low_o_decay}
\end{equation}
In real star, of course, this will not be the case except in or near a
convective zone. Therefore, the field will decay until the regime
$\Omega \gg \omega_\mathrm{A}$ is reached.

\section{The numerical model}

We take advantage of the fact that the Tayler instability is localised
on the meridional plane, and model just a small section of the star on
the rotation axis. This is similar to the arrangement used in Braithwaite 
(2005), where we looked at the Tayler instability in the absence of differential
rotation, modelling only a small volume on the magnetic axis of symmetry.

Inside the computational box, the plasma is rotating in the horizontal
plane about an axis passing through the centre. The rotation speed
$\Omega$ is independent of distance from the axis $\varpi$ and a
function of just height $z$. The bottom face of the box lies in the plane $z=0$. 
The computational box has a height $L$ and width $2L$.

\begin{figure}
\includegraphics[width=1.0\hsize,angle=0]{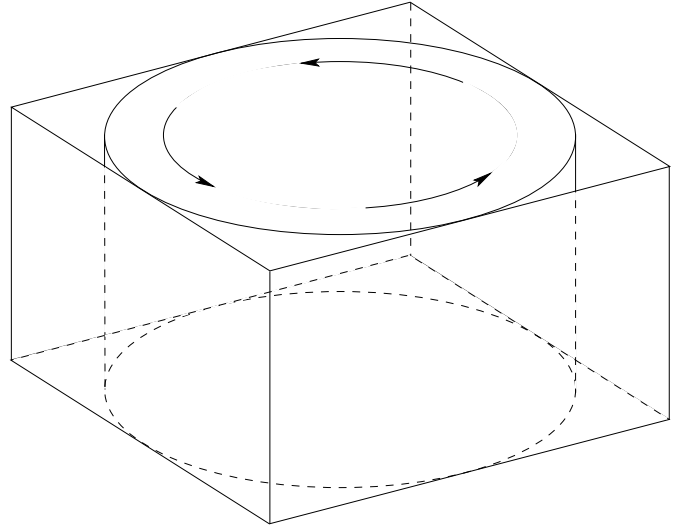}
\caption{The computational box. The cylinder represents the volume
where the gas is being acted on by the rotational force in 
Eq.~(\ref{eq:rotforce}).}
\label{fig:compbox}
\end{figure}

We use a system of Cartesian coordinates. At first glance one might 
instead consider using cylindrical coordinates, as these appear to be 
more suited to the task. This is indeed the case if one is wanting to handle 
the problem analytically. However, cylindrical coordinates not only make 
numerical modelling more time-consuming per grid box, but also introduce 
special points (the axis). This coordinate singularity is a known problem
in all grid-based codes in cylindrical and spherical coordinates.
Since the phenomenon we wish to investigate lies on the z-axis, it is
better to use Cartesians so that we can be sure that our results are
not merely an artefact of the code. The disadvantage is that some
space in the corners of the computational box is wasted. When the
output of the code is analysed, a conversion into cylindrical
coordinates is first performed.
The computational setup is illustrated in Fig.~\ref{fig:compbox}.

\subsection{Implementation of the differential rotation}

We want the rotation rate of the gas to have a time-average dependence on 
height of the form $\Omega=\Omega_0+z {\rm d}\Omega/{\rm d}z$. To achieve
this we apply  a force $\mathbf{F}$ per unit mass of the form:
\begin{equation}
\mathbf{F}(\varpi,z)=(\mathbf{v}_0-\mathbf{v})/\tau_\mathrm{f} \qquad
{\rm where}
\qquad \mathbf{v}_0=\frac{{\rm d}\Omega}{{\rm d}z}z\varpi \mathbf{e}_\phi
\label{eq:rotforce}
\end{equation}
where $\mathbf{v}$ is the velocity field, $\tau_\mathrm{f}$ is a
time-scale, which can be chosen, and $\mathbf{e}_\phi$ is the azimuthal
unit vector. No force was applied in the vertical direction.
The gas at $z=0$ is therefore not rotating. We could in principle add
a constant $\Omega_0$ to the rotation speed, but this would result in
the gas moving more quickly, and the time step of the code would go
down. It is better to include $\Omega_0$ indirectly: we transform to the
rotating frame and add the Coriolis force $2\mathbf{v\times\Omega}_0$
to the momentum equation. The centrifugal force can be ignored since
its only effect would be a change in the equilibrium state.

This force $\mathbf{F}$ is applied to the gas out to a radius $L$, i.e. 
to the sides of the computational box. This means that the corners
escape this force. This was found to be the best way to reduce the
effect of the geometry of the box on the physical phenomenon of
interest -- the gas in the corners is roughly stationary and its effect on
the rotating gas in the middle is minimal.

In a star, differential rotation is driven by two mechanisms: magnetic
braking, which acts on the surface of the star, and evolution, when radial 
shells are contracting and expanding. The former would be difficult to model, 
as exerting a force at the boundaries could be expected to cause problems at 
those boundaries. Exerting a torque by means of a distributed force, as is
done here, is a good way to model differential rotation caused by evolution.

\subsection{Boundary conditions}
\label{sec:bc}


In all three directions, we employ mirror-like boundary
conditions. This means that unknown quantities just outside the boundaries
are copied from the same quantities just inside the same
boundary. This is somewhat more difficult to implement than the
periodic conditions used in Braithwaite (2005), but is necessary for the
following reason. The gas in this case is rotating in a horizontal
plane; with periodic boundary conditions the gas at the boundary would
be moving in the opposite direction to the gas immediately adjacent on
the other side of the boundary. This would cause turbulence -- we are
only interested in one kind of instability and introducing another
would surely confuse the issue. In fact, periodic boundary conditions
were tried at first, and it was found that a dynamo process was
produced even in the case of uniform rotation (${\rm d}\Omega/{\rm d}z=0$), 
as a result of the hydrodynamic turbulence created at the boundaries. This 
argument does not of course hold at the boundaries in the vertical direction,
but there is a good reason for having the same conditions there. Sound
and internal gravity waves can propagate upwards, their amplitude
growing as they do so. With periodic conditions at the vertical
boundaries, a wave can go through the top of the computational
box and come back at the bottom, to continue its upwards
travel. Its amplitude rises exponentially on a timescale roughly equal
to $N^{-1}$ and will eventually get out of control. In Braithwaite 
(2005) we prevented this by having gravity point
in opposite directions in the top and bottom halves of the volume
modelled, the drawback being the computational expense of having to
model the same thing twice. The method used in this study is, as far
the the end result in concerned, the same as that previous method, but
computational cost has been exchanged for programming complexity.

\subsection{Initial conditions}

We begin with uniform temperature; pressure decreases exponentially
with increasing $z$, in hydrostatic equilibrium. The initial velocity
field we create by running the code with no magnetic field until a
steady state has been reached. The initial magnetic field should in
principle be unimportant, except that it must have a vertical
component and must be weak. We choose therefore the simplest field
imaginable: a uniform vertical field $\mathbf{B} = B_0\mathbf{e}_z$,
where $\mathbf{e}_z$ is the vertical unit vector. The field energy
must be weak compared to both the thermal energy and the kinetic
energy, or in other words, the Alfv\'en speed must be much less than
both the sound speed and the rotation speed. $B_0$ is chosen so that
the ratio of thermal to magnetic energy densities $\beta=10^5$, or so
that the ratio of sound to Alfv\'en speeds is $240$. The ratio of
rotation speed to Alfv\'en speed depends of course on the values we
choose for $\Omega_0$ and ${\rm d}\Omega/{\rm d}z$. We want the gas to 
be rotating as fast as possible (to maximize the chances of creating a 
dynamo) but still comfortably below the sound speed. With a magnetic 
field this weak, there is still plenty of space to fit the rotation 
time scale between the sound and Alfv\'en time scales.

\subsection{Free parameters}
\label{sec:free_param}

The goal is to produce a self-sustaining magnetic field, but there are
few clues as to precisely what conditions may be necessary. We have a
fair number of parameters to play with.

The values of $\Omega_0$ and ${\rm d}\Omega/{\rm d}z$ will have an effect. 
It is expected that a value of $\Omega_0$ above $\omega_{\rm A}$ will 
slow the growth rate of the Tayler instability. This would then increase
the saturation field strength. Whether this makes a dynamo any more or
less likely to appear in our model is not clear. What is certain, however, 
is that a large value of ${\rm d}\Omega/{\rm d}z$ will be conducive to
the appearance of a self-sustaining field. We set this therefore in
all runs to a high value but such that the flow speed is still comfortably
below the sound speed. We set ${\rm d}\Omega/{\rm d}z=c_{\rm s}/5L^2$, 
so that the gas is moving at a maximum of one fifth of the sound speed.

Another choice to be made is the relaxation time-scale $\tau_{\rm f}$ of 
the  driving force. Setting it too low would inhibit 
the instability, as it would hold the plasma to too stiff a velocity
field. Too high a value, on the other hand, may mean that the driving
force is insufficient to make the plasma rotate in the required
manner. Various values are used in the results reported below.

The code contains an artificial diffusion scheme designed to maintain
stability.  It includes terms for all three diffusivities (kinetic, 
thermal and magnetic). The adjustable coefficients in this scheme were
set to the experimentally determined minimum value needed for numerical
stability. In the simulations presented in Braithwaite (2005), it was 
possible to turn off this scheme completely, since we were dealing with 
a body of plasma which was stationary at $t=0$ and whose movement we only 
wished to follow while the velocities were small. This is unfortunately 
not the case here, as the plasma is necessarily moving fairly quickly.

In an ideal simulation, all unstable wavenumbers would be modelled between 
the two limits in Eq.~(\ref{eq:minandmaxofn}). The wavenumbers accessible
numerically are given by the vertical size of the computational box 
and the spatial resolution (the Nyquist spatial frequency):
\begin{equation}
\frac{\pi}{{\rm d}z} > n > \frac{2\pi}{L}
\label{eq:num_minandmaxofn}
\end{equation}
To see  an instability  at all, we need this numerical range to overlap 
with  \ref{eq:num_minandmaxofn}. 

\section{The numerical code}

We use a three-dimensional MHD code developed by Nordlund \& Galsgaard
(1995), with extensive modifications, chiefly the mirror-like boundary
conditions described in Sect.~\ref{sec:bc}. The  code uses a staggered
mesh,  so that   variables are  defined   at different  points  in the
grid-box. For example, $\rho$ is defined in the centre of each box, but
$u_x$ in the centre  of the face perpendicular  to the x-axis, so that
the value of $x$ is lower by $\frac{1}{2}{\rm d}x$. Interpolations and
spatial  derivatives  are     calculated to  fifth  and  sixth   order
respectively. The   third order     predictor-corrector  time-stepping
procedure of Hyman (1979) is used. 

The high order of the discretization is  a bit more expensive per grid
point and time  step, but the  code can be run  with fewer grid points
than lower order schemes, for the same accuracy.  Because of the steep
dependence  of computing cost on  grid spacing (4th power for explicit
3D) this results in greater computing economy. 

For stability, high-order diffusive  terms are employed. Explicit  use
is made of   highly   localised diffusivities, while     retaining the
original form of the partial differential equations. 

\section{Results}
\label{sec:results}
We present results for a number of different setups. First, we look at
the unstratified case, both with  and without rotation ($\Omega_0=0$
and $\Omega_0 \ne 0$). Then, the  stratified case, again both with and
without rotation. 

We ran the code with $64$ grid points in the horizontal  directions
and   $32$ in   the   vertical. This is a trade-off between the
resolution needed to obtain dynamo action and the large number of
time steps needed.  The time step  is set by the sound  crossing time, 
but  the (Alfv\'enic) time-scales   of interest to   us are
necessarily much longer. As explained, the high order of the spatial
discretization of the code guarantees a high effective resolution even 
at this limited number of grid points, and self-sustained magnetic 
fields were readily obtained at this resolution in all cases studied.

\subsection{An unstratified case}
\label{sec:unstr}
In the absence of gravity, there is no maximum to the vertical length 
scales that are unstable (see
Eq.~(\ref{eq:minandmaxofn})). Since it is the maximum unstable
wavelength which determines how quickly the field is wound up, the
dynamo never becomes saturated; the field simply continues to grow
until it is strong enough to kill the differential rotation.

However, the fact that we are conducting this simulation inside a box
of finite dimensions changes things somewhat, by imposing an
artificial maximum wavelength. This enables the field to find a
saturation level, and the instability operates at the largest length
scale that fits into the numerical box. 

\begin{figure}
\includegraphics[width=1.0\hsize,angle=0]{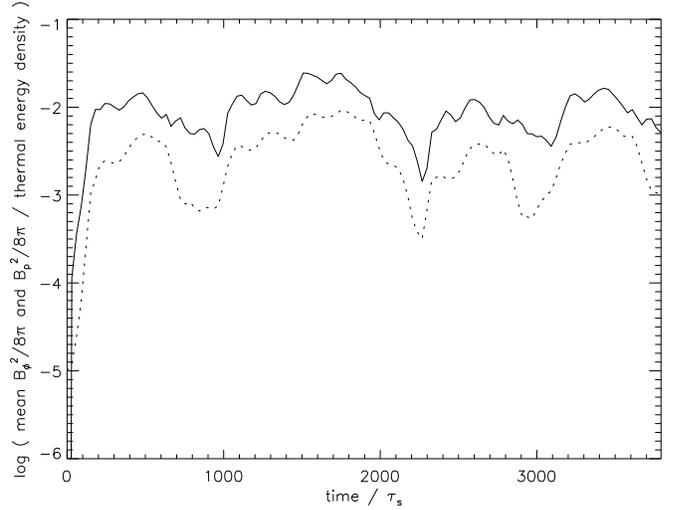}
\caption{Amplitude of the self-sustained magnetic field in an 
unstratified case with maximum shear velocity 1/5 of the sound speed. 
Averages of $B_\phi^2/8\pi$ (solid line) and $B_p^2/8\pi$
(dotted line), in units of the thermal energy density. The field is
predominantly toroidal. Time is in units of $\tau_{\rm s}$, the sound
crossing time of the box.}
\label{fig:me_unstr-props}
\end{figure}

With the unstratified setup, the production of a statistically steady, 
self-sustained magnetic field was observed. To understand the 
properties of the field produced, it is useful to
see the evolution of the mean magnetic energy density,
split up into its poloidal and toroidal components. To this end,
$B_\phi^2/8\pi$ and $B_p^2/8\pi=(B_z^2+B_\varpi^2)/8\pi$ are plotted
in Fig.~\ref{fig:me_unstr-props}. The time on the horizontal axis of
this graph is expressed in units of the sound-crossing time $\tau_{\rm
s}=L/c_{\rm s}$. The field, initially poloidal,
becomes mainly toroidal as it is wound up by the differential
rotation. This happens very quickly, over the time-scale $\tau_{\rm
dr}$. Both components then grow, more slowly, until
the saturation level is reached, when the field is being destroyed by
the instability at the same rate at which it is being amplified by the
differential rotation.

\begin{figure*}
\includegraphics[width=0.333\hsize,angle=0]{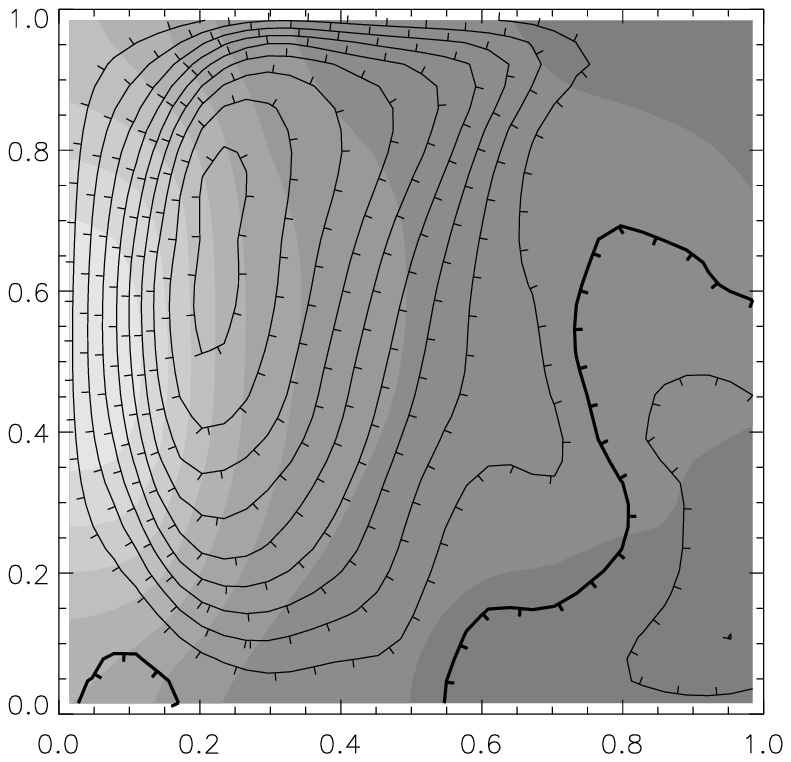}
\includegraphics[width=0.333\hsize,angle=0]{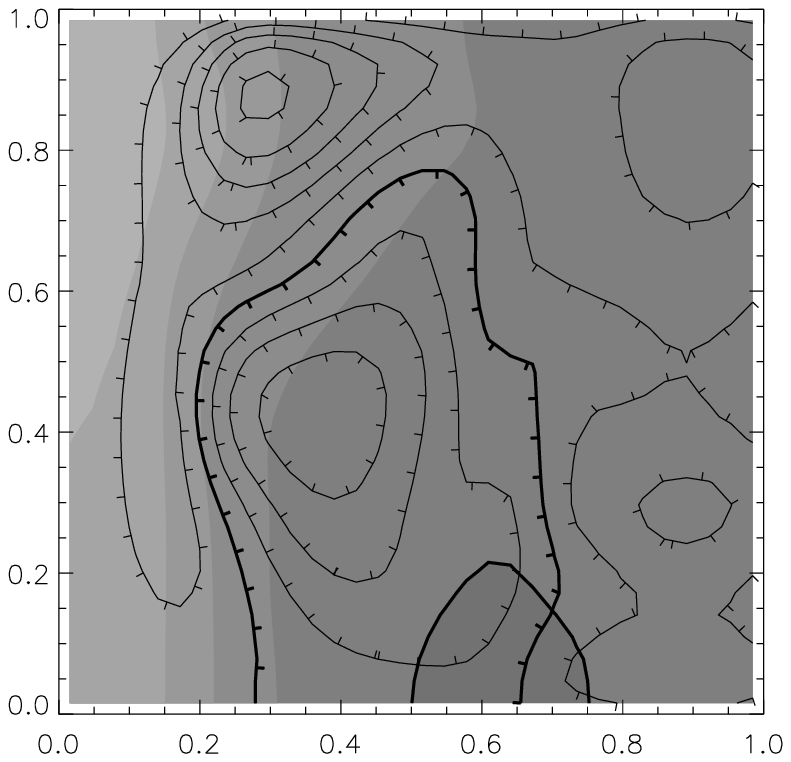}
\includegraphics[width=0.333\hsize,angle=0]{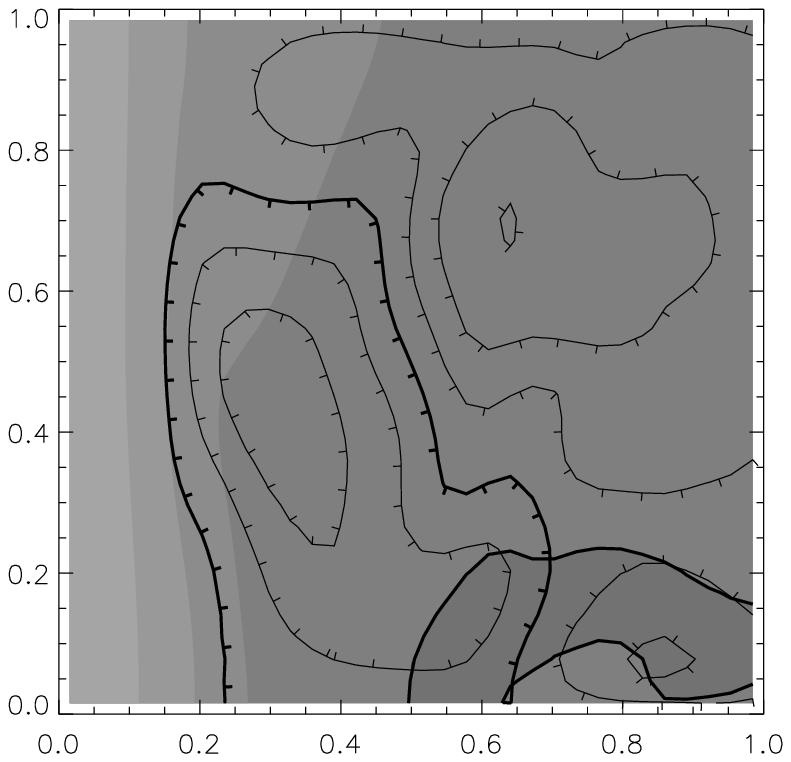}
\\
\includegraphics[width=0.333\hsize,angle=0]{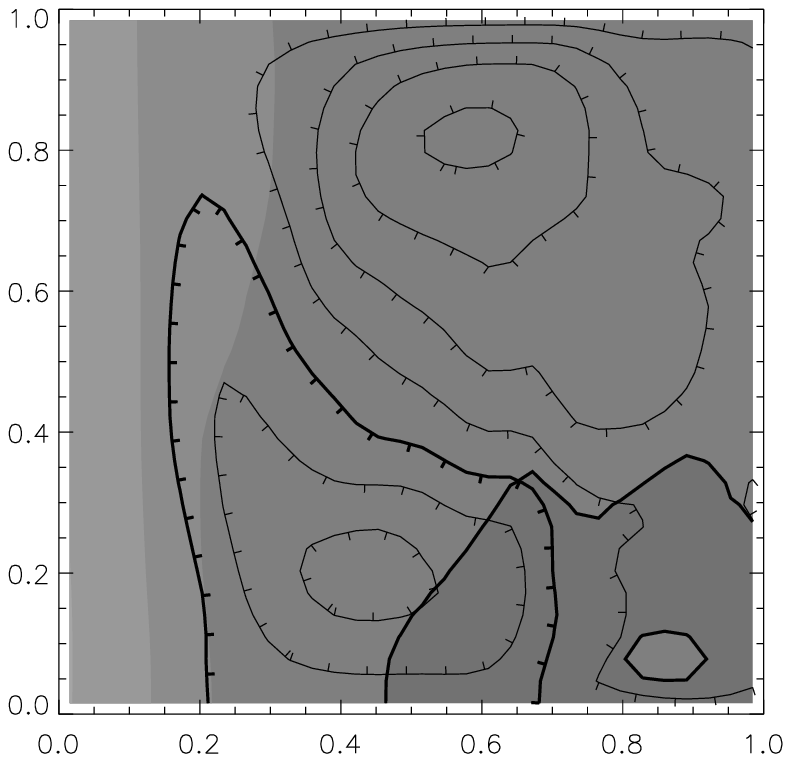}
\includegraphics[width=0.333\hsize,angle=0]{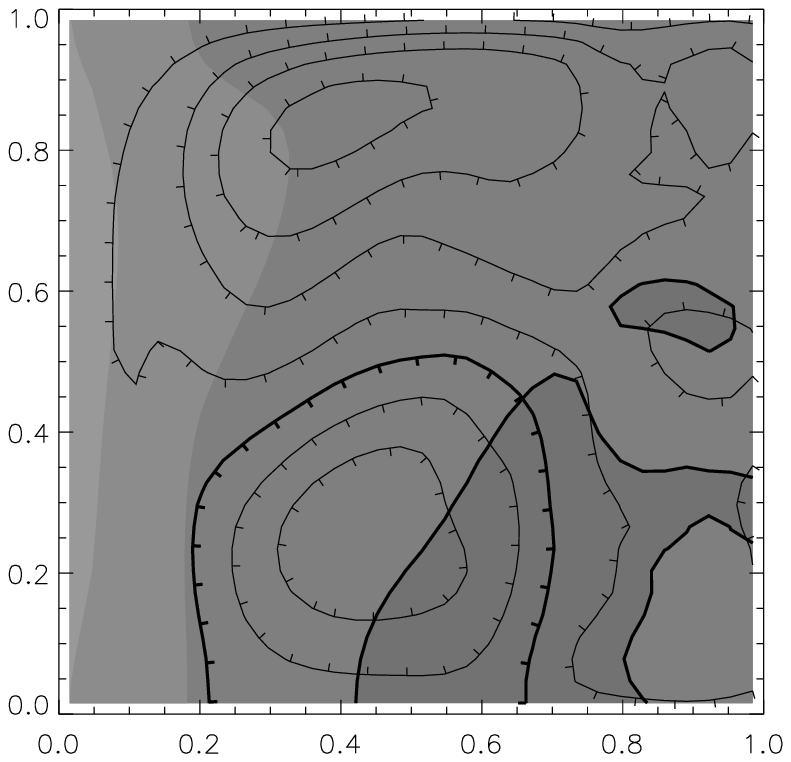}
\includegraphics[width=0.333\hsize,angle=0]{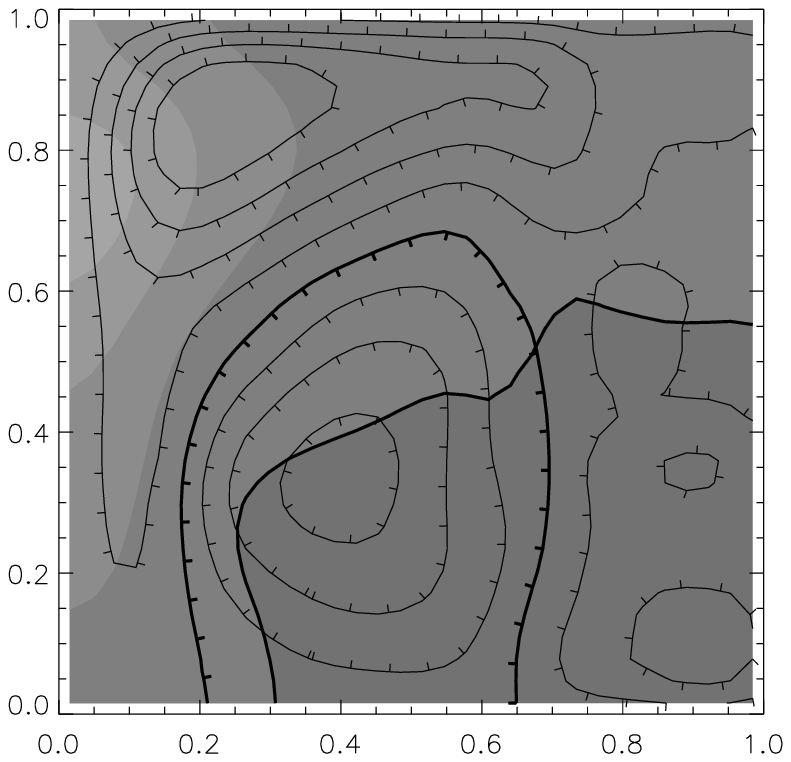}
\\
\includegraphics[width=0.333\hsize,angle=0]{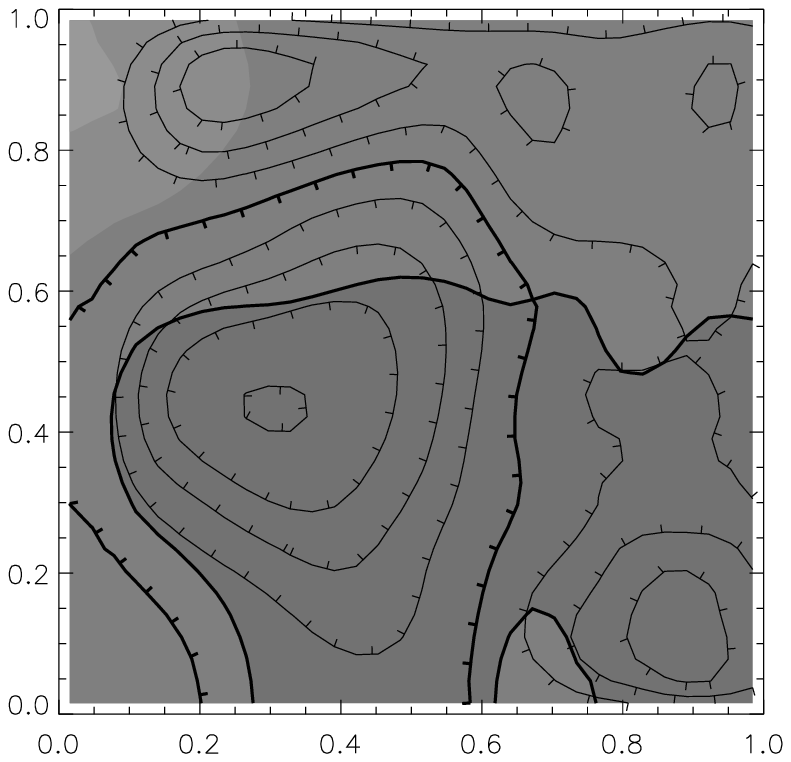}
\includegraphics[width=0.333\hsize,angle=0]{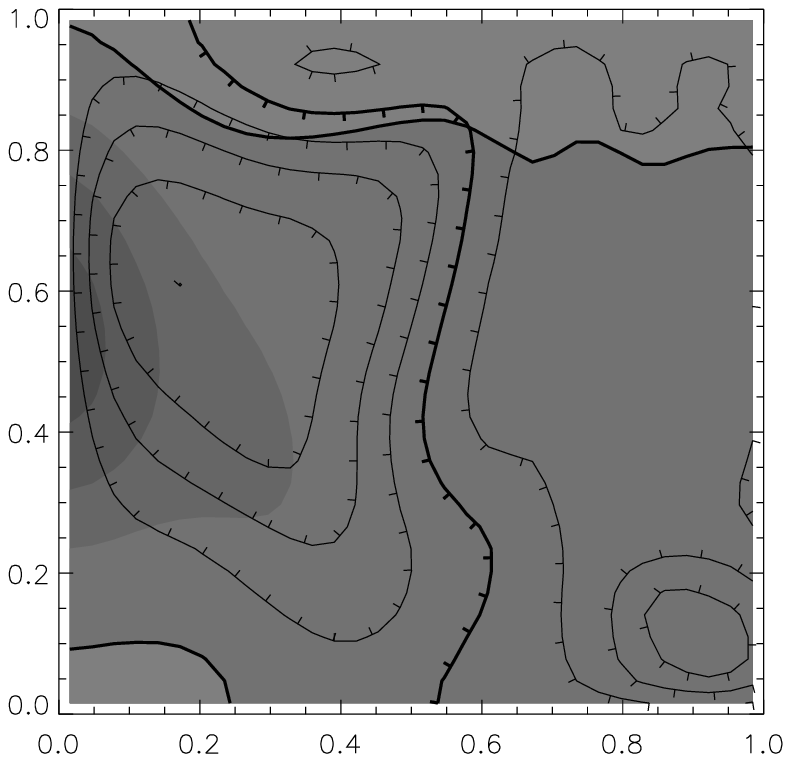}
\includegraphics[width=0.333\hsize,angle=0]{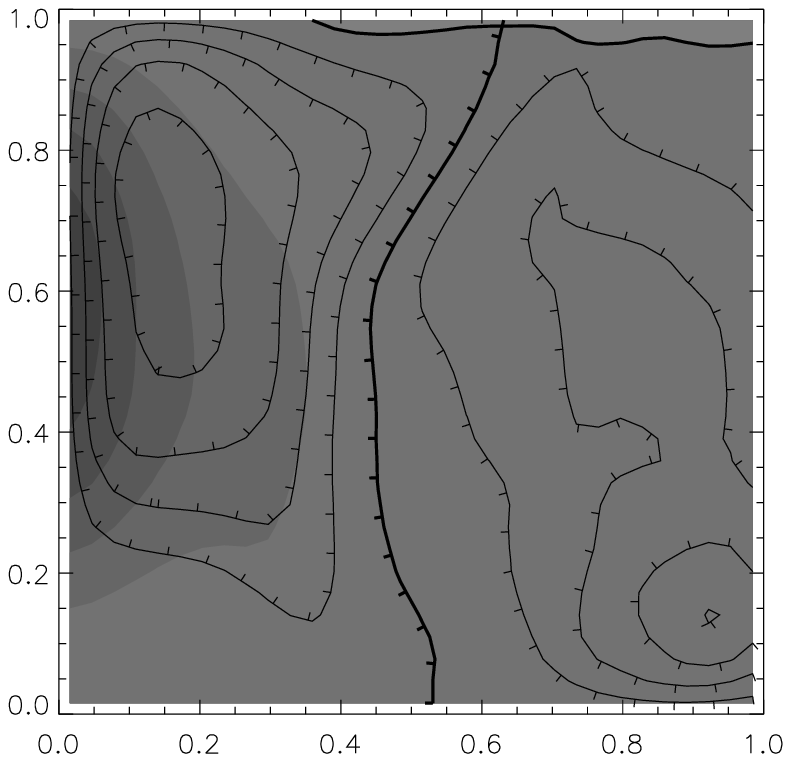}
\caption{Azimuthal average of $B_z$ (positive values shaded light, 
negative dark), as a function of $\varpi$ (horizontal axis) and $z$ 
(vertical axis). Contours show $B_\phi$ (black lines, 
with ticks pointing towards negative values). The thick lines
show where $B_z$ and $B_\phi$ are equal to zero. The nine panels are
taken at nine different times: $t=513$, $664$, $694$, $724$, $754$,
$785$, $815$, $845$ and $875\tau_{\rm s}$, arranged top-left,
top-middle, top-right, middle-left, etc. The predominant sign of
polarities of both $B_z$ and $B_\phi$ changes over this period. Unstratified
case ($N=0$).}
\label{fig:cont-bzandphi_unstr}
\end{figure*}

Fig.~\ref{fig:cont-bzandphi_unstr} shows contour plots of the vertical
component of the magnetic field $B_z$ and of the azimuthal component 
$B_\phi$, averaged in the azimuthal direction, as a function of $\varpi$ 
and $z$. The nine panels are taken at nine different times: $t=513$, 
$664$, $694$, $724$, $754$, $785$, $815$, $845$ and $875\tau_{\rm s}$ 
(in units of the sound crossing time across the box). At the time of the
first frame, the azimuthal mean of $B_z$ is positive everywhere, as it
is at the beginning of the run. $B_\phi$ has been produced from the
winding-up of this positive $B_z$ by differential rotation of positive
${\rm d}\Omega/{\rm d}z$, so is also positive almost everywhere. Then the
instability produces a new vertical component which points
predominantly downwards: $B_z$ switches from positive to negative. The
azimuthal component does likewise.

Looking at Fig.~\ref{fig:me_unstr-props}, it can be seen that the mean
field strength is lower than usual during this reversal period.
This reversal in the prevailing direction of the field happens three
times during this run (also at $t=2200$ and $3000\tau_{\rm s}$), and
there is no reason not to presume that it should continue to
happen were the run continued.

A change of the field direction throughout the entire cylindrical volume 
is not surprising when one
takes into account the fact that only the longest wavelengths are
unstable. Most of the unstable range of wavelengths lies outside of
the range which can be seen in this simulation -- we cannot see
unstable wavelengths longer than the size of the computational box. In
reality, we would see instability over a range of length scales; in
this unstratified case, up to infinity. 

\subsubsection{Torques}
\label{sec:torques}
\begin{figure}
\includegraphics[width=1.0\hsize,angle=0]{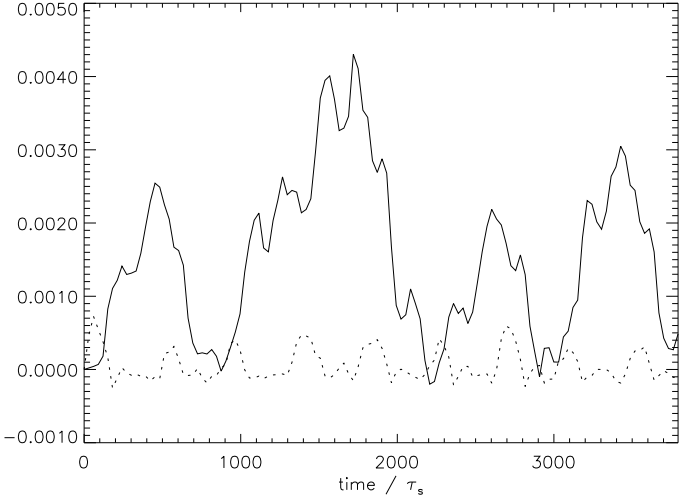}
\caption{Mean torque from the magnetic field $T_{\rm m}$ (solid line) and the
velocity field $T_{\rm v}$ (dotted line), in units of $(2/3)\pi P L^3$. 
Non-stratified ($N=0$), non-rotating ($\Omega_0=0$) case.}
\label{fig:me_unstr-torque}
\end{figure}

The torques produced by the dynamo process vary through the box. 
As an average measure of the magnetic torque, 
we integrate the azimuthal component of the Lorentz
shear stress, multiplied by the lever arm $\varpi$, over a
horizontal plane. We want to check that the torque acting on the
plasma really is magnetic in origin, and not kinetic, since the
velocity field also produces a shear stress. We calculate the two
respective torques in the following way: 
\begin{eqnarray}
T_{\rm m} (z,t) & = & \int_{\phi=0}^{2\pi} \varpi\, {\rm d}\phi
\int_{\varpi=0}^L {\rm d}\varpi\,\varpi \frac{B_\phi B_z}{4\pi},
\label{eq:T_m}\\
T_{\rm v} (z,t) & = & \int_{\phi=0}^{2\pi} \varpi\, {\rm d}\phi
\int_{\varpi=0}^L {\rm d}\varpi\,\varpi \rho v_\phi v_z.
\label{eq:T_v}
\end{eqnarray}
The averages over $z$ of these two torques are plotted in
Fig.~\ref{fig:me_unstr-torque}. This confirms that the torque is
chiefly magnetic. The torque from the velocity field is almost ten
time smaller, sometimes even negative, i.e. it has the effect of 
{\it increasing} the differential rotation.

The torques $T_{\rm m}$ and $T_{\rm v}$ are plotted in units of
$(2/3)\pi PL^3$; this is the torque which would exist if the shear 
stress were equal to the gas pressure $P$. In accretion disk parlance, 
this corresponds to a viscosity paramater $\alpha=1$. In
fact, it is of some interest to compare the shear stress in this
model to that found in an accretion disc, since they are caused by
similar processes. That the stress in an accretion disc is of the
order of the gas pressure should not be surprising, since the kinetic
energy of the orbital motion, which drives the dynamo, is of
the same order as the thermal energy. In a differentially rotating
star, however, the energy available to power the dynamo (the 
difference in kinetic energy compared with a uniformly rotating star 
with the  same angular momentum), is much less than the thermal energy. 
We should therefore also expect the magnetic energy to be much less 
than the thermal energy, as it cannot be greater than the energy of 
differential rotation. 

\begin{figure}
\includegraphics[width=1.0\hsize,angle=0]{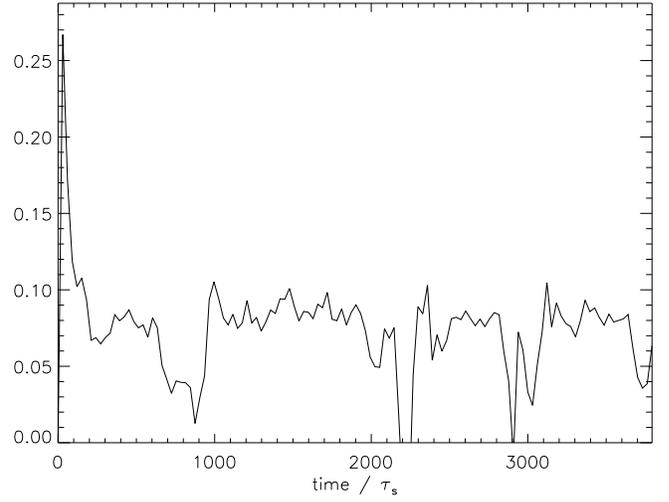}
\caption{Efficiency coefficient $\epsilon$, as defined in
Eq.~(\ref{eq:epsilon}) for the run shown in Fig. \ref{fig:me_unstr-torque}}
\label{fig:unstr-epsilon}
\end{figure}

We have seen that the torque exerted by the magnetic field is much
less than the torque which would exist if the shear stress were equal
to the gas pressure, as we expected. A more interesting quantity to
compare it to, in this context, is the torque which would exist if the
shear stress were equal to the {\it magnetic} pressure $B^2/8\pi$. To
do this, we calculate a dimensionless efficiency coefficient
$\epsilon$, defined such that:
\begin{equation}
T_{\rm m}(z,t) = \epsilon(z,t) \int_{\phi=0}^{2\pi} \varpi\, {\rm
d}\phi \int_{\varpi=0}^L {\rm d}\varpi\,
\varpi \frac{B^2}{8\pi}.
\label{eq:epsilon}
\end{equation}
This efficiency $\epsilon$, or rather, the average of it over all $z$,
is plotted is Fig.~\ref{fig:unstr-epsilon}. After the initial
winding-up phase, its value tends to stay at around $0.08$, except for
the field-reversal phases when the field is weaker.

If the vertical and azimuthal component of the field were everywhere
equal, and if the component in the $\varpi$ direction were zero, we
would have $\epsilon=1$. As in the case of MRI turbulence in accretion 
disks, however, the field is mainly azimuthal.
Comparing Eqs.~(\ref{eq:T_m}) and (\ref{eq:epsilon}), and assuming
that the ratio $B_\phi/B_z$ is the same everywhere and that
$B_\varpi=0$:
\begin{equation}
\epsilon \approx \frac{B_\phi B_z/4\pi}{B^2/8\pi} \approx
\frac{2(B_\phi/B_z)}{1+(B_\phi/B_z)^2}.
\label{eq:epsilon-ratios}
\end{equation}
Looking at Fig.~\ref{fig:me_unstr-props}, we can estimate that
$B_\phi/B_z \approx 1.8$, and the above equation gives us $\epsilon
\approx 0.8$. The main reason for the low torque efficiency may then
be that the ratio $B_\phi/B_z$ is not constant, rather, it is low at
small $\varpi$ and high at large $\varpi$. This is confirmed by
looking at the last frame of Fig.~\ref{fig:cont-bzandphi_unstr}, for
instance: $B_z$ is strongest very close to the axis, as the
Tayler instability is strongest there, and $B_\phi$ is strongest somewhat
further from the axis, because the winding-up effect is stronger at
larger $\varpi$.

\subsubsection{Parameter dependence}
For the run discussed above the net rotation $\Omega_0=0$, (the rotation 
rate at $z=0$, the middle of the box), and the damping time of the applied 
force $\tau_{\rm f}$ was set equal to the sound crossing time $\tau_{\rm s}$. 
For values of $\tau_{\rm f}$ much different from this, a self-sustaining 
field was not produced. If too low a value is used, the velocity field is 
too `stiff' and the instability is unable to take hold, although the field
does reach the required strength. If $\tau_\mathrm{f}$ is too high, on
the other hand, the differential rotation is slowed down by the
Lorentz forces from the magnetic field before the necessary field
strength for instability has been reached.

\begin{figure}
\includegraphics[width=1.0\hsize,angle=0]{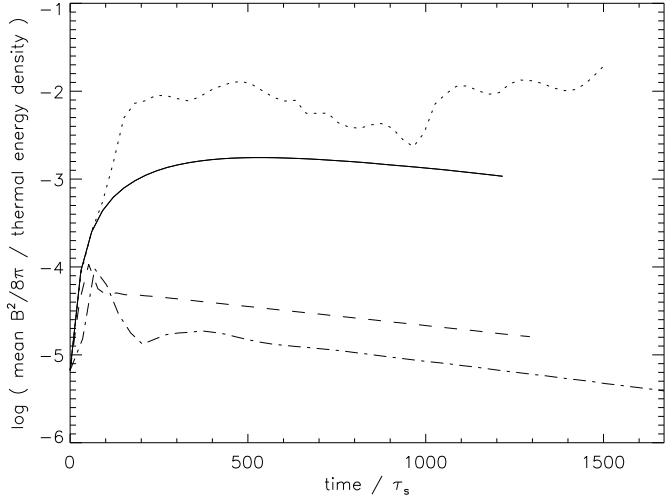}
\caption{Mean magnetic energy density against time, for runs with 
$N=0$ and $\tau_\mathrm{f}/\tau_{\rm s}=$ $0.1$ (solid line),
$1.0$ (dotted), $10$ (dashed) and $100$ (dot-dashed). Energy density 
is in units of the thermal energy density, time in sound crossing times 
$\tau_{\rm s}$. Only in a narrow range of $\tau_\mathrm{f}/\tau_{\rm s}$
is a statistically steady magnetic field maintained.}
\label{fig:me_unstr-tf}
\end{figure}

The code was run with a range of values of $\tau_\mathrm{f}/\tau_{\rm s}$: 
$0.1$, $1.0$, $10$ and $100$. Fig.~\ref{fig:me_unstr-tf}.
shows the evolution of the mean magnetic energy density $B^2/8\pi$.
It can be seen that a self-sustaining field was observed only around
$\tau_\mathrm{f}/\tau_{\rm s}=1$.
It was also found that the dynamo is not produced if a lower spatial
resolution is used. It therefore seems that where the dynamo {\it was}
produced, the conditions were only just sufficient. So it should not be
surprising that it also only works in a narrow range of the parameter 
$\tau_\mathrm{f}$. A higher spatial resolution will be needed to obtain 
dynamo action in a less restricted range of parameter values. 

\subsubsection{Rotation}
In the above we have shown a working dynamo in simulations with 
differential rotation, but
with no rotation overall. This is of course unlikely to be the
situation inside a real star, so we shall now model a plasma with a net
rotation by adding a Coriolis force to the momentum equation.
We find that rotation above a certain speed is able to stop the
dynamo from working, presumably because with the given resolution it
is only just possible to see the instability working, and small
changes can push the dynamo action just outside the effective range. We
shall see in the next section that in the stratified case the dynamo
is more vigorous; active in a less restricted range of parameters.

\begin{figure}
\includegraphics[width=1.0\hsize,angle=0]{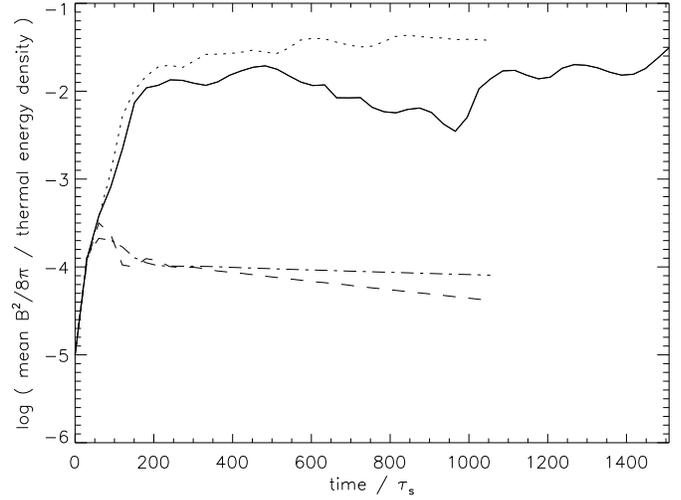}
\caption{Mean magnetic energy density against time, for $N=0$ runs with 
forcing parameter $\Omega_0/\tau_{\rm s}^{-1}=$ $0$ 
(solid line), $0.1$ (dotted), $0.3$ (dashed) and $1.0$ (dot-dashed). Energy 
density in units of the mean thermal energy density. Sustained magnetic 
fields are obtained only at the lower rotation rates.}
\label{fig:me_unstr-omega}
\end{figure}

We run the code as above, but with values of $\Omega_0$ which will put 
us into the $\Omega \gg \omega_\mathrm{A}$ regime:
$\Omega_0/\tau_{\rm s}^{-1}=0.1$, $0.3$ and $1.0$.
Fig.~\ref{fig:me_unstr-omega} shows the magnetic energy
density as a function of time, for these runs, as well as for the
$\Omega_0=0$ run. It can be seen that when the rotation speed is above
some certain threshold, the field, although still wound up at first,
decays again without reaching
a steady state. This behavior is not fully understood. It is expected
that rotation will slow the growth of the instability, and will
therefore increase the minimum unstable wavelength
(Eq.~(\ref{eq:minandmaxofn})), which depends on diffusivity. This could
mean that the field never becomes unstable.

\subsection{The stratified case}
\label{sec:str}
In the stratified case, the additional parameter is the buoyancy frequency 
$N$. The main result 
will turn out to be that a self-sustaining magnetic field is much
easier to create in the stratified case. Whereas in the unstratified
case, a self-sustained field was only produced within a narrow range of
parameters, with stratification the range of parameters is now much
wider. Rotation does not destroy the dynamo, even for rotation frequency
$\Omega$ as large as the buoyancy frequency.

There are one or two other differences. The field builds up to saturation 
much more slowly -- in the unstratified case, saturation was reached over 
one or two Alfv\'en crossing times (Alfv\'en crossing times at the initial 
field strength). In the stratified case, it takes at least five times
longer. Also, the field energy is much steadier; the fluctuations do
not seem to be as large.

\begin{figure}
\includegraphics[width=1.0\hsize,angle=0]{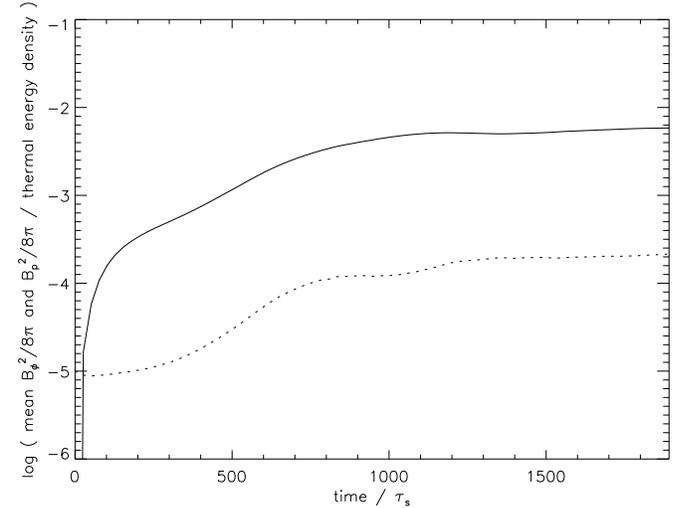}
\caption{The means of $B_\phi^2/8\pi$ (solid line) and $B_p^2/8\pi$
(dotted line), in units of the thermal energy density, for a stratified
case $N=\tau_{\rm s}^{-1}$. The field is even more predominantly toroidal 
than in the unstratified case (compare Fig.~\ref{fig:me_unstr-props}).}
\label{fig:me_str-props}
\end{figure}

Fig.~\ref{fig:me_str-props} shows the toroidal and poloidal components
of the magnetic energy, for a run with $N=\tau_{\rm s}^{-1}$ and
$\tau_{\rm f}=10\tau_{\rm s}$. The energy in the toroidal component is 
around $30$ times larger
than that in the poloidal component, a larger difference than in the
absence of stratification (cf. Fig.~\ref{fig:me_unstr-props}).

\begin{figure}
\includegraphics[width=1.0\hsize,angle=0]{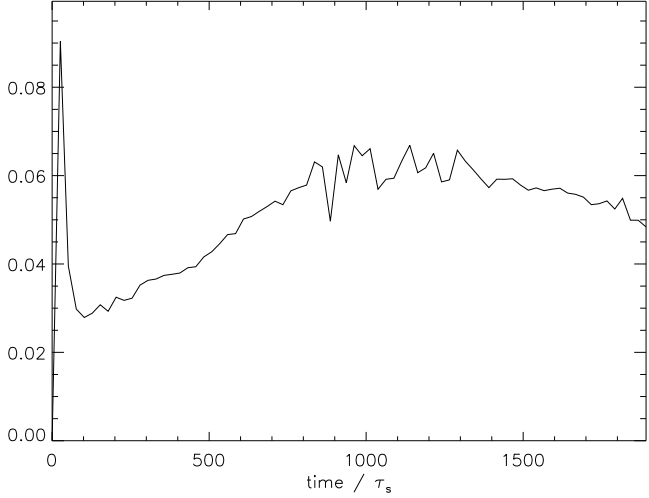}
\caption{Efficiency coefficient $\epsilon$, as defined in
Eq.~(\ref{eq:epsilon}). Stratified ($N=\tau_{\rm s}^{-1}$),
non-rotating case.}
\label{fig:str-epsilon}
\end{figure}

As in Sect.~\ref{sec:torques}, we calculate an efficiency coefficient
$\epsilon$ (see Eq.~(\ref{eq:epsilon})). This is plotted in
Fig.~\ref{fig:str-epsilon} for this stratified run. The value settles
a little lower than in the unstratified case, at around
$0.05$. In this case, however, the ratio $B_\phi/B_z$ is much greater
(comparing Figs.~\ref{fig:me_unstr-props} and \ref{fig:me_str-props}),
so we expect a lower efficiency $\epsilon$. The reason that the same
value of $\epsilon$ is measured despite a higher average of the ratio
$B_\phi/B_z$ must have something to do with the respective
distributions as a function of $\varpi$ of the vertical and azimuthal
components of the field.

We can try using other values of the driving-force time-scale
$\tau_{\rm f}$, to see how robust the dynamo is.
For these runs, we used a somewhat stronger initial magnetic field,
with $\beta=10^3$, as opposed to $\beta=10^5$ used in the previous
runs. This helps to speed up the initial evolution a little, but has
no effect on the final steady state.

\begin{figure}
\includegraphics[width=1.0\hsize,angle=0]{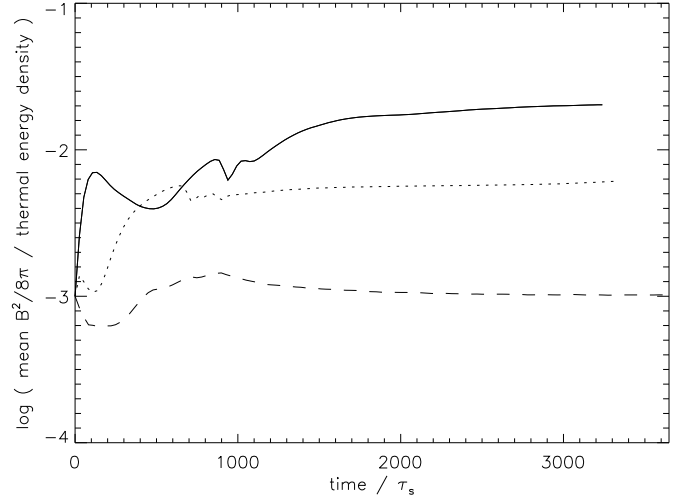}
\caption{Mean magnetic energy density against time, for the stratified
case $N=\tau_{\rm s}^{-1}$.  $\tau_{\rm f}/\tau_{\rm s}=$ $1$ (solid line), 
$10$ (dotted) and $100$ (dashed). A self-exciting field is produced 
in all three cases.}\label{fig:me_str-tf}
\end{figure}

We use values of $\tau_{\rm f}/\tau_{\rm s}$ of $1$, $10$ and
$100$. In this last case, the time-scale $\tau_{\rm f}$ is much
greater than other relevant time-scales, which will also be the case
in a real star. The magnetic energy in these runs is plotted in
Fig.~\ref{fig:me_str-tf}. A self-sustaining field appears in all
three cases, although the saturation field strength does depend to
some extent on the value of $\tau_{\rm f}$. This in fact does reflect
reality -- the faster the differential rotation is being driven, the
stronger we expect the excited field to be.

\begin{figure}
\includegraphics[width=1.0\hsize,angle=0]{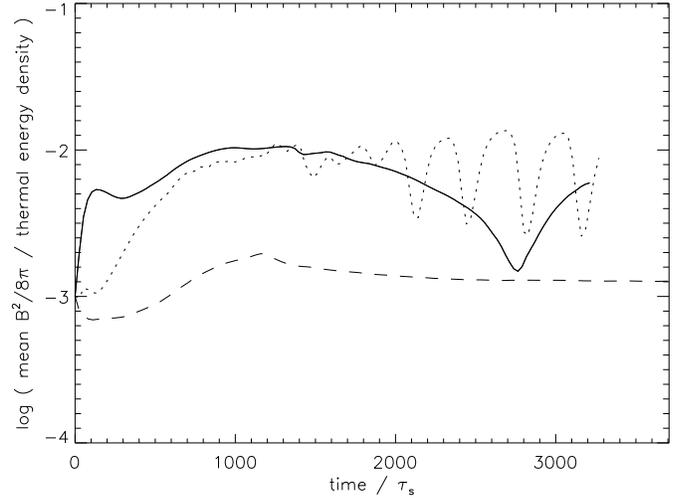}
\caption{Mean magnetic energy density against time, for a rotating, 
stratified case. 
$\tau_{\rm f}/\tau_{\rm s}=$ $1$ (solid line), $10$ (dotted) and $100$
(dashed). A self-exciting field is produced in all three cases, and a
oscillatory behavior is seen in one case.}
\label{fig:me_str-omega-tf}
\end{figure}

We have now looked at three cases: non-rotating unstratified;
rotating unstratified; and non-rotating stratified. The fourth
combination -- rotating stratified ($\Omega\gg\omega_{\rm A}$ and
$N\gg\omega_{\rm A}$) -- will exist in almost the entire
non-convective zone of almost all stars, and is therefore the most
interesting to us. To be sure that we are in this regime, we shall set
both $N$ and $\Omega_0$ to the reciprocal of the sound-crossing time,
i.e. to $\tau_{\rm s}^{-1}$.

It is found that the self-sustaining field is still produced, unlike
when rotation was added to the unstratified case. The field produced
is of comparable strength to that produced when rotation is absent,
but oscillates in a seemingly regular fashion. The mean values of
$B_\phi$ and $B_z$ go from positive to negative, and back again. We
saw something like this in Sect.~\ref{sec:unstr}, but there, the
reversal of the field was a more chaotic process with, one presumes,
no regular period.  Fig.~\ref{fig:me_str-omega-tf} is the rotating
equivalent of Fig.~\ref{fig:me_str-tf}. Except at the highest value of
$\tau_{\rm f}$, the field energy can be seen to jump up and down in a
regular way.


\section{Discussion and Conclusions}

We have demonstrated by direct numerical simulation a dynamo process as 
envisaged by analytical arguments in Spruit (2002). It feeds off differential 
rotation {\it only}, and not from any imposed small-scale velocity field. 
The toroidal (azimuthal) component of the field is produced from the
winding-up of the existing poloidal (meridional) component. This toroidal 
field is unstable, and produces, as a result of its decay, a new poloidal
component, which can then be wound up itself -- in this way, the
`dynamo loop' is closed. A weak seed field is amplified in this way
until a certain saturation level is reached, at which the field is
being wound up by differential rotation at the same rate as it is
decaying through its inherent magnetohydrodynamic instability.

This dynamo is therefore different from the traditional convection-powered 
dynamo models for stars with convective envelopes, in which the dynamo
loop is closed by  poloidal components induced by an additional, 
{\it non-magnetic} process. The dynamo found here is, in this sense, 
similar to the MRI turbulence in accretion disk (\cite{Hawleyetal:1996}) 
which is powered by the gradient in orbital rotation in the disk.

A dominant factor in the case of differential rotation in a star is the 
effect of the stable stratification. It strongly restricts the types of 
instability that can occur in the azimuthal field produced by winding up
in the differential rotation. In Spruit (1999) we have shown that the
first instability to set in as the magnetic field strength increases by this
process is Tayler-instability (a pinch-type instability) rather than 
magnetic buoyancy instabilities. 

In the simulations presented here we have considered separately the effects
of rapid net rotation and a stabilizing stratification. Perhaps somewhat
surprisingly, dynamo action was found to set in more readily (at lower spatial
resolution) in the stratified cases than in the unstratified case. This is
the opposite of what would be expected if buoyancy instabilities were the
dominant mechanism closing the dynamo loop. This confirms the conclusion in 
Spruit (2002) that differential rotation in a stable stratification leads 
to a self-sustained magnetic field in which Tayler instability of the 
azimuthal field is the dominant process closing the dynamo cycle.

The results presented were obtained only at a resolution close to the 
minimum required to obtain dynamo action. As a result, the field configurations
obtained show only minimal structure in the vertical direction. Effectively,
they cover only vertical length comparable with the characteristic vertical
length scale of the process (which is governed by the strength of the
stratification, cf. Spruit 2002). 

For further progress, simulations at higher resolution will allow more
detailed comparison with the analytical estimates. An important step
forward, however, would be the development of code more suited to the
nearly-incompressible, highly stratified conditions relevant for stellar
interiors. With the existing codes, the time step limitations due to sound 
speed and the buoyancy frequency limit the degree of realism that can be 
achieved.

\begin{appendix}

\end{appendix}


\begin{flushleft}

\end{flushleft}

\end{document}